\documentclass{article}

\usepackage{arxiv}
\usepackage[utf8]{inputenc}
\usepackage{amssymb,amsmath,latexsym}
\usepackage{soul,color}
\usepackage{xcolor}
\usepackage{graphicx,amssymb,amsmath}
\usepackage{algorithm}
\usepackage[noend]{algpseudocode}
\usepackage{pgfplots}
\usepackage{array}
\usepackage[utf8]{inputenc}
\newcolumntype{P}[1]{>{\centering\arraybackslash}p{#1}}
\usepackage{tikz}
\usepackage{caption}
\usepackage{pgfplots}
\usepackage{hyperref}
\usepackage{float}
\usepackage{subcaption}
\pgfplotsset{compat=newest}
\usepackage[utf8]{inputenc}
\usepackage{hyperref}
\usepackage{array}
\usepackage{graphicx}
\usepackage{multirow}

\usepackage[outdir=./]{epstopdf}   
\usepackage{dblfloatfix}
\usepackage[figuresright]{rotating}

\title{Enhancing Neural Spoken Language Recognition: An Exploration with Multilingual Datasets}

\usepackage{authblk}

\newbox{\orcid}\sbox{\orcid}{\includegraphics[scale=0.06]{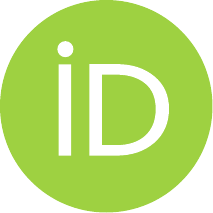}} 
\author[3,1,4]{%
	\href{https://orcid.org/0000-0002-7673-6511}{\usebox{\orcid}\hspace{1mm}Or Haim Anidjar\thanks{\texttt{orchaimanidjar@gmail.com }}}%
}

\author[1,2]{%
	\href{https://orcid.org/0000-0001-5180-4541}{\usebox{\orcid}\hspace{1mm}Roi  Yozevitch\thanks{\texttt{yozevitch@gmail.com}}}%
}
\affil[1]{Department of Computer Software \& Engineering, Ariel University, Golan Heights 1, 4077625, Ariel, Israel}
\affil[2]{Department of Electrical Engineering, Ariel University, Golan Heights 1, 4077625, Ariel, Israel}
\affil[3]{Faculty of Computer Science, College of Management, Elie Wiesel 2, 7579806, Rishon LeTsiyon, Israel}
\affil[4]{Ariel Cyber Innovation Center, Ariel University, Golan Heights 1, 4077625, Ariel, Israel}

\hypersetup{
pdftitle={nhancing Neural Spoken Language Recognition: An Exploration with Multilingual Datasets},
pdfsubject={ASR},
pdfauthor={Or Haim Anidjar, Roi Yozevitch},
}

\begin{document}
\maketitle

\begin{abstract}

In this research, we advanced a spoken language recognition system, moving beyond traditional feature vector-based models. Our improvements focused on effectively capturing language characteristics over extended periods using a specialized pooling layer. We utilized a broad dataset range from Common-Voice, targeting ten languages across Indo-European, Semitic, and East Asian families. The major innovation involved optimizing the architecture of Time Delay Neural Networks. We introduced additional layers and restructured these networks into a funnel shape, enhancing their ability to process complex linguistic patterns. A rigorous grid search determined the optimal settings for these networks, significantly boosting their efficiency in language pattern recognition from audio samples. The model underwent extensive training, including a phase with augmented data, to refine its capabilities. The culmination of these efforts is a highly accurate system, achieving a 97\% accuracy rate in language recognition. This advancement represents a notable contribution to artificial intelligence, specifically in improving the accuracy and efficiency of language processing systems, a critical aspect in the engineering of advanced speech recognition technologies.

\end{abstract}



\section{Introduction}
\label{sec:intro}

Over the years, the field of spoken language recognition (SLR) has developed remarkably, becoming both a fascinating and crucial area of study. The history of SLR can be traced back to early rule-based systems, followed by statistical and machine learning-based approaches in the 1980s and 1990s, respectively. In recent years, the field has seen the successful application of advanced statistical models like GMMs, as well as deep learning-based approaches such as CNNs and LSTM networks \cite{wu2017introduction,zhao2016machine,basnin2020integrated}.

The importance of SLR is underscored by its diverse applications, ranging from security systems that detect unauthorized access to customer service and virtual assistants. It also plays a crucial role in language acquisition, providing feedback on pronunciation and intonation, and in speech therapy, where it can diagnose speech abnormalities and offer tailored treatment. These applications make SLR an indispensable tool in modern society.

The major purpose of SLR research is to develop technology and systems for the automatic and more accurate identification, analysis, and understanding of spoken language. This is particularly vital in an era where communication crosses the borders of countries and cultures \cite{siniscalchi2013universal}. As technology advances, SLR systems are likely to become even more accurate and robust, paving the way for increased cross-cultural communication and breaking down language barriers.

The advancement of SLR technology holds enormous implications for cross-cultural communication. It has the potential to create greater understanding and collaboration among individuals from various backgrounds, thereby playing a critical role in breaking down language barriers.

SLR encompasses a range of techniques designed to identify and analyze spoken language from audio data. These techniques fall into two main categories: feature-based and model-based methods \cite{adaloglou2020comprehensive}. Feature-based methods focus on extracting specific acoustic features, such as pitch, tone, accent, and speed, to characterize the audio signal. In contrast, model-based methods develop specialized models to capture language-specific characteristics. Commonly used techniques in SLR include the Gaussian Mixture Model (GMM), Hidden Markov Model (HMM) \cite{eddy2011accelerated}, Convolutional Neural Network (CNN) \cite{mukherjee2019spoken}, and Long Short-Term Memory (LSTM) \cite{zhang2020graph}. The choice of technique often depends on the specific application and the amount of available training data.

Within this framework, a specialized area known as speaker embedding aims to create compact and unique numerical representations of individual voices \cite{stafylakis2019self}. This task can be viewed as an extension of both feature-based and model-based methods. The x-vector technique, for example, employs a Deep Neural Network (DNN) to extract a variable-length numerical representation of a voice, capturing both short-term acoustic~\cite{anidjar2023stethoscope} features like pitch and tone, as well as long-term language patterns \cite{snyder2018x}. The i-vector approach, on the other hand, uses a factor analysis model combined with a Gaussian Mixture Model-Universal Background Model (GMM-UBM) to generate a fixed-length numerical representation, primarily capturing speaker-specific characteristics \cite{morrison2011comparison}. The d-vector method is similar to the x-vector but is trained on a more extensive dataset and longer speech sequences, making it robust across varying speech conditions \cite{radfar2020end}. Both x-vector and d-vector techniques utilize neural networks, aligning them closely with model-based methods and making them adept at handling complex feature spaces and temporal fluctuations in voice signals.  While these techniques have significantly advanced the field of SLR, they are not without their limitations and challenges.

\subsection{Challenges in SLR Research}
Despite the significant advancements in spoken language recognition, the field still faces several challenges that hinder the production of accurate and dependable results \cite{sahu2018challenges}. One of the most pressing issues is background noise, which can originate from various sources such as traffic, wind, or even other speakers in the vicinity. This noise can significantly reduce the accuracy of SLR systems~\cite{anidjar2024whisper, anidjar2024harnessing}.

Another major hurdle is the variability in dialects and accents. Even within the same language, people often speak with different accents and dialects, complicating the task for voice recognition algorithms. This challenge is particularly pronounced in linguistically diverse countries like India, where each region may have its own unique dialect and accent.

The scarcity of high-quality training data~\cite{anidjar2024extending} poses an additional challenge. Speech recognition algorithms heavily rely on training data to improve their accuracy over time. However, obtaining large volumes of quality data can be problematic, especially for less commonly spoken languages or dialects. This often results in under-trained models that lack both accuracy and reliability.

Real-time processing of spoken language adds another layer of complexity. SLR systems often struggle to identify language in real-time, particularly during spontaneous speech, such as genuine conversations. The need for fast and efficient algorithms capable of handling massive data volumes in real-time further complicates this issue.

These challenges underscore the critical need for ongoing research and development in SLR. To enhance the performance of these systems, technological breakthroughs like improved algorithms and innovative data collection methods are essential.

In light of these challenges, our work aims to introduce innovative solutions that enhance the accuracy and reliability of SLR systems.

\subsection{Our Contribution}

To address the existing challenges in SLR, the primary contribution of this paper is the exploration and application of the Vector-X technique. This novel approach allows us to effectively represent various aspects of speech, including words, grammatical structure, and specific speech processing like tone, in a vector space. These representations serve as a robust and consistent feature set for machine learning models aimed at recognizing and classifying spoken language. Furthermore, we introduce methods to enhance the performance of x-vector and other embedding models through different architectures and various data augmentations. Remarkably, we achieve enhanced accuracy in Time Delay Neural Network (TDNN) and x-vector algorithms while utilizing less data compared to baseline models. Our algorithm underwent rigorous training and testing on three major language families—Semitic, Indo-European, and East Asian—and ten sub-languages. The selection of these languages is strategic, aiming to address the global language barrier as they represent a significant portion of non-English speakers worldwide.

\subsection{Paper Structure}

The remainder of this paper is structured as follows: 
Section 2 presents previous works that mainly used x-vector and TDNN and different methods; 
Section 3 discusses the dataset used in the article, how it is built and what it consists of how it was chosen, and the pre-processing; 
Section 4 presents the framework employed in the article and the modification made to the new TDNN model; 
Section 5 shows the process of the experiment and the results of the new model including the modification and exploitation of the architectural structure that is different from the basic model; 
Finally, Sections 6 and 7 provide a discussion, conclusions, and suggestions for future follow-up work.
For ease of reading, Table 2 provides a list of abbreviations that are commonly used in this paper.

\begin{table}[th]
  \caption{List of abbreviations}
  \label{tab:list_of_abbreviations}
  \centering
  \begin{tabular}{ |c|c|c|}
  \hline
    \multicolumn{1}{|c|}{\textbf{Abbreviation}} & 
    \multicolumn{1}{|c|}{\textbf{\# Meaning}}\\\hline
    ASR & \quad Automatic Speech Recognition \\\hline
    CNN & \quad  Convolutional Neural Network  \\\hline
    DNN & \quad Deep Neural Network  \\\hline
    DNN & \quad Depp Neural Network  \\\hline
    GMM & \quad Gaussian mixture models  \\\hline
    GPU & \quad Graphics Processing Unit  \\\hline
    HMM & \quad  Hidden Markov Model  \\\hline
    LID & \quad language identification  \\\hline
    LSTM  & \quad Long Short-Term Memory Networks \\\hline
    MMI & \quad  Maximum Mutual Information \\\hline
    NIST & \quad National Institute of Standards Technology \\\hline
    NN & \quad Neural Network  \\\hline
    RNN & \quad Recurrent Neural Network  \\\hline
    SLR & \quad Spoken language recognition \\\hline
    SVM & \quad Support Vector Machine  \\\hline
    TDNN  & \quad Time delay neural network \\\hline
    UBM & \quad Universal Background Model  \\\hline
    WAV & \quad Waveform Audio File Format  \\\hline
  \end{tabular}
\end{table}

\section{Related Work}

Spoken language recognition (SLR) has been a topic of significant interest in the speech processing community. Various techniques and methodologies have been proposed to enhance the accuracy and robustness of SLR systems.

\subsection{Feature Extraction Techniques}
The extraction of meaningful features from speech signals is a crucial step in SLR. Traditional systems have relied on i-vectors for this purpose. The conventional method for i-vector extraction utilizes a Universal Background Model (UBM) and employs a projection matrix $T$, which is trained to enhance the likelihood of the training data. \cite{matvejka2011full}. Recent advancements have introduced x-vectors and d-vectors as alternatives to i-vectors. Table 1 provides a comparison of these techniques. As observed, i-vectors utilize a low-dimensional representation and are based on a generative modeling approach, typically employing a Gaussian Mixture Model-Universal Background Model (GMM-UBM) architecture. In contrast, x-vectors use a high-dimensional representation, adopt a discriminative modeling approach, and are often associated with Convolutional Neural Network (CNN) architectures. D-vectors, on the other hand, combine features of both, resulting in a hybrid modeling approach. They also use a high-dimensional representation and can be trained using both supervised and unsupervised methods, typically leveraging Long Short-Term Memory (LSTM) architectures \cite{wang2019discriminative}.
\begin{table}[H]
\centering
\caption{Comparison of i-vector, x-vector, and d-vector between Technique by the Representation, Modeling, Training, Architecture}
\label{tab:vector-comparison}
\resizebox{0.5\textwidth}{!}{%
\begin{tabular}{|c|c|c|c|}
\hline
\textbf{Technique} & \textbf{i-vector} & \textbf{x-vector} & \textbf{d-vector} \\
\hline
Representation & Low-dim. & High-dim. & High-dim. \\
\hline
Modeling & Generative & Discriminative & Hybrid \\
\hline
Training & Supervised & Supervised & Sup./Unsup. \\
\hline
Architecture & GMM-UBM & CNN & LSTM \\
\hline
\end{tabular}%
}
\end{table}

In a study by \cite{alan2014multiclass}, i-vectors were trained using a Multiclass Discriminative approach. \cite{alan2014multiclass} created a technique in which they trained a Gaussian i-vector classifier using Maximum Mutual Information (MMI). This method was designed to optimize the multiclass calibration criterion, eliminating the need for a separate back-end. When tested on the NIST LRE11 standard evaluation, the results indicated that this streamlined, single-stage approach maintained high performance and calibration. Furthermore, the system was adapted for open set tasks by incorporating the additive Gaussian noise model, which underwent discriminative training to enhance its performance.

X-vectors have emerged as a powerful representation of SLR. Contrary to i-vectors that concentrate on low-dimensional representations, x-vectors utilize deep neural networks to identify more complex patterns in speech data. The versatility of x-vectors allows them to represent various aspects of speech, including linguistic content, speaker attributes, and even emotional tone. Recent studies, such as \cite{snyder2018spoken} have highlighted the effectiveness of x-vectors in SLR, especially when combined with data augmentation and discriminative Gaussian classifiers.

\subsection{Deep Learning Approaches}
Deep neural networks (DNNs) have shown promise in enhancing SLR systems. DNNs, when trained as acoustic models for Automatic Speech Recognition (ASR), have been incorporated into i-vector systems for example~\cite{rouvier2014speaker}. Another study, titled "Generative Adversarial Networks for Noise-robust Language Identification" \cite{8639522}, leveraged generative adversarial nets (GAN) combined with DNN-based i-vector approaches for language identification. Furthermore, the application of deep convolutional recurrent neural networks has been explored for speech emotion recognition in specific languages, as discussed in \cite{swain2022dcrnn}

\subsection{Multilingual and Dataset Considerations}
Training models on multilingual datasets can enhance their generalization capabilities. One study, titled "Multilingual Joint Training of Sequence-to-sequence Models for Spoken Language Identification" \cite{alan2014multiclass}, trained a model on nine different Indian languages and found improved recognition performance. Another research effort investigated the use of web audio data collected from YouTube for $107$ languages, emphasizing the importance of post-filtering to ensure data quality\cite{valk2021voxlingua107}.

\subsection{Enhancements and Optimizations}
Various optimizations have been proposed to further improve SLR systems. For instance, the use of conditional generative adversarial nets (cGAN) has been explored for optimizing parameters, as discussed in "Generative Adversarial Networks for Noise-robust Language Identification" \cite{8639522}. Another study, \cite{daniel2016stacked}, introduced a time delay neural network (TDNN) for SLR, achieving significant gains.

\subsection{X vectors}

Contemporary language recognition systems predominantly depend on i-vectors. The standard approach employs a UBM and a substantial projection matrix $T$, both trained to optimize the likelihood of the training data \cite{matvejka2011full}. This method transforms the high-dimensional statistics of UBM into a compact, low-dimensional i-vector. Post-extraction, i-vectors are classified using Gaussian models, logistic regression, or cosine scoring. Advanced i-vector systems incorporate deep neural networks (DNN) trained as acoustic models for automated speech recognition (ASR) \cite{rouvier2014speaker}, often in one of two ways: replacing Gaussian mixture model (GMM) posteriors with ASR DNN posteriors, or training the i-vector system with bottleneck features derived from the DNN.

In their study, \cite{alan2014multiclass} trained i-vectors with Multiclass Discriminative, demonstrating a method where a Gaussian i-vector classifier is trained using Maximum Mutual Information (MMI) to directly optimize the multiclass calibration criterion, eliminating the need for a separate back-end. This approach, confirmed by results on the NIST LRE11 standard evaluation task, maintains high performance and calibration with this streamlined methodology. Additionally, the system extends to the open set task using a discriminatively trained additive Gaussian noise model to enhance performance.

\cite{daniel2016stacked} explored a novel approach by introducing a stacked architecture employing a time delay neural network (TDNN) for modeling long-term patterns in spoken language identification. The architecture consists of a feed-forward neural network with a bottleneck layer trained to classify context-dependent phone states (senones), followed by a TDNN that processes the bottleneck output over an extended time span to generate language probabilities. This system surpasses a state-of-the-art shifted delta cepstra i-vector system and offers complementary insights for integration with new bottleneck-based i-vector systems. It achieves a relative 27\% increase in identification accuracy.

\cite{snyder2018spoken} applied x-vectors to spoken language recognition and DNN tasks. They discovered that a temporal pooling layer in the network captures long-term language characteristics by aggregating information over time. X-vectors, once extracted, utilize the same classification technologies as i-vectors. Their best-performing system incorporates multilingual bottleneck features, data augmentation, and a discriminative Gaussian classifier.

Lastly, \cite{Jorgen2020VOXLINGUA107} investigated leveraging web audio data collected automatically. They generated search phrases from language-specific Wikipedia data to retrieve YouTube videos in 107 languages. Techniques like speech activity detection and speaker diarization were used to isolate speech segments from these videos. The resultant training set, VoxLingua107, amounts to 6628 hours (averaging 62 hours per language), supplemented by a 1609-utterance evaluation set. They also implemented post-filtering to enhance the accuracy of language labeling in the database, achieving a 98\% correctness rate based on crowd-sourced verification.

\section{Datasets}

In the field of speech recognition, the ability to accurately transcribe and understand diverse languages is of utmost importance. To address this challenge, our work utilized datasets sourced from Common Voice, a valuable resource for multilingual speech data \cite{ardila2019common}. We specifically focused on ten languages, carefully selected from three distinct language families: Indo-European, Semitic, and East Asian. This article aims to shed light on the significance of recognizing and supporting speech recognition in these languages, considering the substantial number of non-English speakers worldwide \cite{hunt2004self}. We believe that with these languages Our work aimed to contribute to the development of multilingual speech recognition systems. Such systems have the potential to bridge language barriers, empower non-English speakers, and foster inclusivity on a global scale.

\subsection{Indo-European Languages}

Among the Indo-European language family, we chose to include Spanish, French, Italian, and Russian in our study. These languages were selected for their remarkable diversity and widespread usage across multiple continents. By incorporating languages from different regions, we aimed to capture the richness and variation present within this language family \cite{fortson2011indo}. Detailed information about each of these languages, including the number of recorded hours, validated hours, and the number of different speakers, is presented in Table \ref{table:Indo-European-languages}.

\begin{table}[h]
\centering
\caption{Details of Languages in the Indo-European Family Used in the Study}
\label{table:Indo-European-languages}
\begin{tabular}{|c|c|c|c|}
\hline
Language & Recorded Hours & Validated Hours & Number of Speakers \\
\hline
Spanish  & 24            & 11              & 207                \\
\hline
French   & 12            & 15              & 298                \\
\hline
Italian  & 4             & 4               & 49                 \\
\hline
Russian  & 8             & 5               & 103                \\
\hline
\end{tabular}
\end{table}

\subsection{Semitic Languages}

n our research, we recognized the importance of Semitic languages, which have historical and cultural significance in various parts of the world \cite{hetzron2009semitic}. To represent this language family, we focused on Arabic, Hausa, and Farsi. These languages are spoken by millions of people globally and play a vital role in their respective communities. By including Semitic languages in our study, we aimed to address the specific challenges and intricacies associated with their speech recognition. Detailed statistics about each of these languages, including the number of recorded hours, validated hours, and the number of different speakers, are presented in Table \ref{table:Semitic-languages}.

\begin{table}[h]
\centering
\caption{Statistics for Semitic Languages}
\label{table:Semitic-languages}
\begin{tabular}{|c|c|c|c|}
\hline
Language & Recorded Hours & Validated Hours & Number of Speakers \\
\hline
Arabic & 2 & 1 & 77 \\
\hline
Hausa & 13 & 4 & 39 \\
\hline
Farsi & 2 & 4 & 37 \\
\hline
\end{tabular}
\end{table}

\subsection{East Asian Languages}

The third group we explored consisted of East Asian languages. Chinese, Japanese, and Thai were selected as representatives of this diverse and vibrant language family. East Asian languages possess unique phonetic structures and tonal characteristics, making them intriguing subjects for speech recognition research. Given the sizable populations that speak these languages, their accurate recognition holds great significance for enabling effective communication and access to technology for non-English speakers \cite{henderson1965topography}. Detailed statistics for each language in the East Asian language family, including the number of recorded hours, validated hours, and the number of different speakers, are presented in Table \ref{table:East-Asian-languages}.

\begin{table}[h]
\centering
\caption{Statistics for East Asian Languages}
\label{table:East-Asian-languages}
\begin{tabular}{|c|c|c|c|}
\hline
Language & Recorded Hours & Validated Hours & Number of Speakers \\
\hline
Chinese & 5 & 95 & 189 \\
\hline
Japanese & 4 & 3 & 36 \\
\hline
Thai & 7 & 5 & 77 \\
\hline
\end{tabular}
\end{table}

It's important to note that our dataset includes both male and female speakers across various age groups and dialects. However, specific distribution statistics for each language were not available.

The Common Voice dataset distinguishes between Recorded and Validated hours. Recorded hours represent the total duration of submitted voice recordings, while Validated hours are those that have been manually reviewed and approved by the community for quality criteria such as clarity and naturalness \cite{lee2008principles}.

\subsection{Modeling Objectives and Their Influence on Dataset Selection}
Our model prioritizes understanding languages over individual speakers for several reasons:

\begin{itemize}
    \item \textbf{Generalization}: Ensuring universal recognition irrespective of the speaker for real-world efficacy \cite{lee2008principles}.
    \item \textbf{Variability}: A diverse set of speakers in training data guarantees robustness against unique speech attributes.
    \item \textbf{Privacy}: Emphasizing language over speaker identity enhances privacy.
    \item \textbf{Scalability}: Given the vast number of potential speakers, focusing on languages is more efficient.
    \item \textbf{Data Imbalance}: Addressing potential biases from imbalanced speaker representation \cite{garcia2016stacked}.
\end{itemize}

\subsection{Dataset Pre-processing and Validation}
The Common Voice dataset differentiates between Recorded hours, the total duration of submitted recordings, and Validated hours, which have undergone community review for quality. Our research evaluated the necessity of pre-processing on the Common Voice dataset. Although considered clean, we converted the mp3 files to wav format for compatibility and consistency reasons. Furthermore, we removed "dead" segments, or portions with minimal voice activity, to enhance data accuracy \cite{maughan2019comparing}.

\subsection{Balanced Data and Cleaning}
We balanced our dataset by selecting an equal number of voice segments from each language, mitigating potential biases from imbalanced representation. This approach ensures unbiased analysis and improved model generalization \cite{mirus2020importance}.

\subsection{Data Augmentation}
Data augmentation enhances dataset diversity and robustness. Adhering to methodologies in \cite{park2019specaugment}, we employed several techniques:

\begin{itemize}
    \item \textbf{Speed Variation}: Modulating audio clip speed to simulate speech rate variations among speakers \cite{ko2015audio}.
    \item \textbf{Pitch Perturbation}: Shifting audio signal pitch to capture human voice variability \cite{hasija2021out}.
    \item \textbf{Noise Addition}: Introducing various types of noise, including white, pink, and environmental, to prepare the model for real-world noisy scenarios \cite{park2019specaugment}.
\end{itemize}

\section{Framework}

\subsection{Baseline Model and New Model}

\begin{figure}[htp] 
    \centering
    \includegraphics[width=8cm]{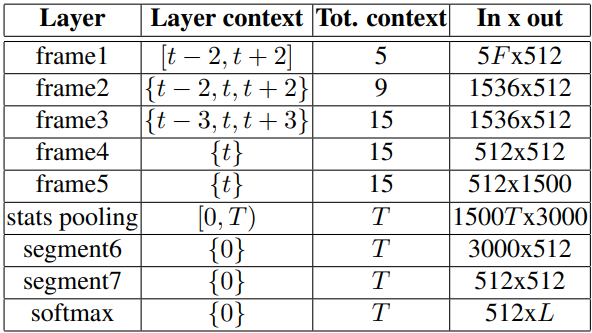}
    \caption{Architecture Source Model (Baseline) based on the \href{github.com/KrishnaDN/x-vector-pytorch}{Spoken Language Recognition using X-vectors}.} \label{tab:arc_source}
\end{figure}

This subsection compares the baseline Deep Neural Network (DNN) architecture with the new model proposed in this paper. The baseline DNN, as illustrated in Figure~\ref{tab:arc_source}, processes an input sequence of \(T\) speech frames with initial layers maintaining a limited temporal context. The statistical pooling layer aggregates data across temporal dimensions, resulting in a 3000-dimensional vector fed through segment-level layers to the softmax output layer, as detailed in Figure~\ref{tab:arc_source}.

The enhanced architecture of our research model, presented in Table~\ref{tab:model}, introduces two additional intermediate layers and adopts a funnel pattern in subsequent layers. This revised structure aims to optimize generalization and performance.

\begin{table}[ht]
\centering
\caption{An enhanced architecture of our model, characterized by modifications from the baseline. This includes the incorporation of additional layers and adjustments to the input-output values}
\label{tab:model}
{%
\begin{tabular}{|c|c|c|c|c|}
\hline
\textbf{Layer} & \textbf{Layer context} & \textbf{Tot. context}  & \textbf{Tot. width of context} & \textbf{In x out}\\
\hline
frame 1 & \{t-2,t+2\}& 3 & 5 & 3Fx1280 \\
\hline
frame 2 & \{t\} & 3 & 5 & 1280x1280 \\
\hline
frame 3 & \{t-4,t-2,t,t+2,t+4\} & 7 & 9 & 8960x1024 \\
\hline
frame 4 & \{t\} & 7 & 9 & 1024x1024 \\
\hline
frame 5 & \{t-1,t+1\} & 9 & 11 & 9216x768 \\
\hline
frame 6 & \{t\} & 9 & 11 & 768x512 \\
\hline
frame 7 & \{t\} & 9 & 11 & 512x256 \\
\hline
stats pooling & [0,T) & T & T & 256Tx512 \\
\hline
segment 8 & \{0\} & T & T & 512x512 \\
\hline
segment 9 & \{0\} & T & T & 512x512 \\
\hline
softmax & \{0\} & T & T & 512xL \\
\hline
\end{tabular}%
} 
\end{table}

Key modifications to the Baseline Model include:
\begin{enumerate}
    \item Conducting a Grid Search on 'context size' and 'dilation' in TDNN layers \cite{probst2018tunability}.
    \item Incorporating intermediate layers for enhanced frame and context processing \cite{al2019character}.
    \item Transitioning to a funnel configuration in TDNN layers \cite{yu2007importance}.
\end{enumerate}

Each modification, discussed in detail in subsequent sections, contributes to the integrated model's improved performance, which will be presented in the results section.

\begin{figure}[htp]
    \centering
    \includegraphics[width=6.7cm]{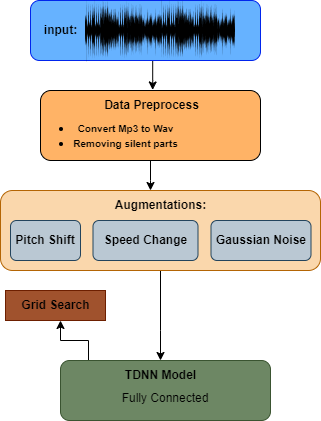}
\caption{Network topology: The initial phase involves inputting an mp3 file, which undergoes transformation to wav format during data preprocessing. This is followed by augmentations like pitch shift, speed modulation, and Gaussian noise addition, before feeding into the TDNN model, looping over a Grid Search.}
\end{figure}

\subsection{Architecture improvements}

\subsubsection{Context Size and Dilation TDNN}

In Temporal Convolutional Neural Network (TDNN) layers, optimizing context size and dilation is essential for effectively modeling sequential data, particularly in speech recognition tasks \cite{demirel2020automatic}. Context size, the number of adjacent time frames considered during convolution, determines the range of temporal information captured, with larger sizes enabling the incorporation of a broader temporal context crucial for understanding long-term dependencies \cite{yu2020densely}. Dilation, on the other hand, refers to the spacing between frames in the convolution process and is key to capturing non-adjacent temporal relationships, thus aiding in modeling long-range dependencies without markedly increasing computational load. The optimization of these parameters, a critical aspect of TDNN layers, demands a balance between capturing necessary temporal dependencies and avoiding computational inefficiency or overfitting. In this context, grid search proves invaluable. Our refined approach goes beyond basic grid search methodologies, which typically iterate over a preset range of values, by tailoring the search range with task-specific insights. This targeted approach focuses on optimizing context size and dilation specifically for speech recognition, adding an extra layer of specificity and effectiveness to the optimization process.

A grid search was conducted to optimize the dilation and context size hyperparameters in TDNN layers, a systematic method ensuring thorough exploration of all possible hyperparameter combinations \cite{madaan2017wire}. This exhaustive approach, albeit computationally demanding, was crucial for determining the most effective settings for our model. The search was conducted separately for each of the first three layers due to computational constraints, with a limited number of epochs per layer to expedite the process. The optimal hyperparameters identified were:

\begin{itemize}
    \item Layer 1 with a context size of 3 and dilation of 2,  achieving an accuracy of 0.72 and a loss of 0.88.
    \item Layer 2 with a context size of 5 and dilation of 2, resulting in an accuracy of 0.66 and a loss of 1.07.
    \item Layer 3 with a context size of 2 and dilation of 1, leading to an accuracy of 0.85 and a loss of 0.49.
\end{itemize}

It's important to note that these values, while indicating a direction for model improvement, may not be the absolute optimal due to the limited epoch count and the restricted range of tested hyperparameter values.

\subsubsection{1x1 TDNN Intermediate Layers}

In our research, we integrated 1x1 TDNN intermediate layers into the baseline model to enhance its performance \cite{snyder2018spoken}. 1x1 TDNN layers, characterized by a context size and dilation of 1, are specialized temporal convolutional layers that process sequential data such as speech by focusing exclusively on the current frame or time step. This approach is effective in capturing immediate temporal relationships and fine-grained local patterns in the data \cite{toledano2018multi,li2018improving}.

These layers introduce non-linear transformations to the input data, making them particularly valuable for tasks requiring detailed analysis of the current time step. In speech recognition and language processing, 1x1 TDNN layers can significantly enhance the model's ability to detect subtle temporal patterns at a granular level.

In our model, we added two specific 1x1 TDNN layers, \textit{tdnn2} and \textit{tdnn4}, positioned as intermediate layers. These layers, taking outputs from tdnn1 and tdnn3 respectively, focus solely on the current frame, enhancing feature transformation and increasing the model's non-linearity. By doing so, they help in capturing complex patterns in the input more effectively. Their inclusion is vital for increasing the model's capacity to discern intricate patterns, thereby boosting its discriminative power for tasks like language detection.

These intermediate layers contribute to the model's overall capability by providing additional non-linear mappings and refining the representations of input features. This is particularly beneficial for tasks that require a nuanced understanding of both global and local language patterns, such as language detection, where capturing the specific linguistic characteristics at various abstraction levels is crucial for accurate classification.

\subsubsection{TDNN Funnel Structure}

Our research modifies the TDNN network structure from the original configuration of five layers with 512 neurons each to a funnel-shaped, or ``bottleneck,'' architecture, as discussed in the literature \cite{tishby2015deep}. This funnel structure offers several advantages:

\begin{enumerate}
    \item \textbf{Dimensionality Reduction}: The funnel structure progressively reduces data dimensionality, leading to efficient and compact representations.
    \item \textbf{Hierarchical Feature Extraction}: It enables hierarchical processing, where each layer abstracts higher-level features from the data, aiding in learning complex patterns.
    \item \textbf{Information Bottleneck and Regularization}: The narrowing structure acts as an information bottleneck, prioritizing critical information and preventing overfitting by focusing on the most informative features for improved model performance.
    \item \textbf{Efficient Computation}: The reduced dimensionality in deeper layers decreases computational requirements, resulting in faster training and inference, beneficial for large datasets and constrained resources.
    \item \textbf{Parameter Efficiency and Robustness}: The network becomes more parameter-efficient and robust to noise and variations in the input data, with initial layers providing a global view and subsequent layers refining localized information.
    \item \textbf{Adaptation to Data Size}: Given the specific data size in our research, the layer sizes were adapted differently from the suggested Fourth root of the data size, considering the unique requirements of our dataset and model.
\end{enumerate}

This TDNN funnel structure is particularly relevant to our research focus, enhancing the network's ability to process and learn from complex data patterns effectively.

\subsection{Integrated Model and Augmentations}
In our final model, we combined three key enhancements: values optimized through grid search, 1x1 TDNN intermediate layers, and a funnel structure. This integrated approach significantly strengthened the model's capabilities, as detailed in the previous sections. To further refine its performance, the model was trained on augmented data, which included adjustments in audio speed, added noise, and altered pitch. This augmentation aimed to enhance the model's robustness and unreliability, exposing it to a wider range of realistic audio conditions. As a result of these combined improvements, the model achieved an impressive accuracy of $95\%$ on the validation set, demonstrating its effectiveness in diverse and challenging audio environments.

\section{Experimental Evaluation and Results}

In accordance with the methodologies outlined in the framework, several modifications were independently evaluated. Below is a tabulated comparison illustrating the differences among these variations:

\begin{table}[ht]
\centering
\caption{Comparative Analysis of Model Improvements on Validation Dataset.}
\label{tab:results_comparison}
\resizebox{0.4\textwidth}{!}{%
\begin{tabular}{|c|c|}
\hline
\textbf{Model} & \textbf{Accuracy} \\
\hline
Baseline & 0.54 \\ 
\hline
Grid Search & 0.85 \\ 
\hline
Intermediate Layers & 0.79  \\ 
\hline
Funnel Structure & 0.92\\ 
\hline
Integrated Approach & 0.95  \\ 
\hline
Final Model & 0.969  \\ 
\hline
\end{tabular}%
}
\end{table}

Baseline Model: The initial approach was based on the model described in Snyder et al. (2018), utilizing the x-vector method as implemented in the available GitHub repository: "https://github.com/KrishnaDN/x-vector-pytorch". Given the lack of a predefined dataset, a flexible choice of datasets was necessary. After selecting an appropriate dataset from the Common Voice database and training the model with default parameters in a Google Colab environment, the model achieved an accuracy of 0.54.

Grid Search: As an initial enhancement, a grid search was conducted to identify optimal model parameters (refer to the framework's grid-search subsection). This process yielded an improved accuracy of 0.85.

1x1 TDNN Intermediate Layers: The addition of two intermediate layers resulted in an accuracy increase of 0.25. Further details on the impact of these layers can be found in the dedicated section of the paper.

Funnel Structure in TDNN Layers: Implementing a funnel structure in the TDNN layers, starting with 1280 neurons and decreasing by 256 neurons per layer, resulted in a notable performance boost, achieving an accuracy of 0.92.

Integrated Model: This model combined the previous three enhancements (1x1 TDNN intermediate layers, Grid Search, Funnel Structure). The integrated approach demonstrated promising results, achieving an accuracy of 0.95.

Final Model: Incorporating all the aforementioned improvements, the final model achieved maximal accuracy. Further enhancement was achieved through data augmentation, including the addition of background noise, variable audio speeds, and pitch alterations, which contributed significantly to the model's robustness and accuracy in test scenarios.

Figures depicting the confusion matrices for the final model on both validation and test datasets are included, demonstrating the model's effectiveness and potential overfitting issues, respectively.

\begin{figure}[htp]
    \centering
    \includegraphics[width=14.0cm]{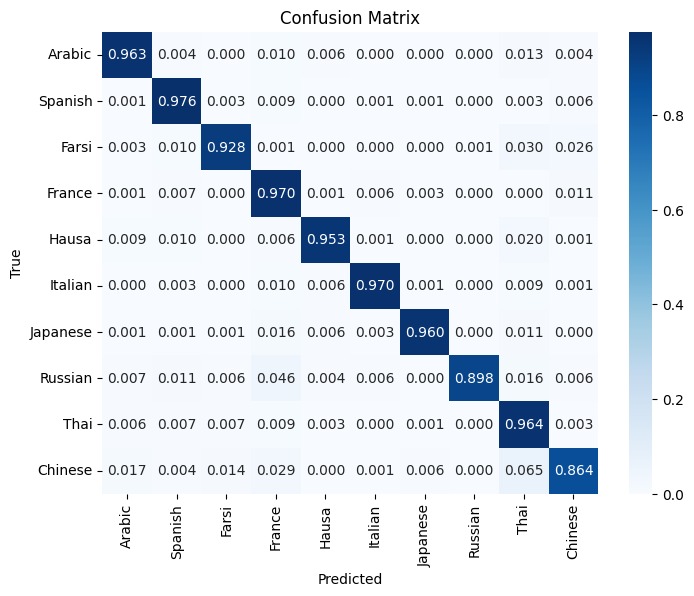}
    \caption{Confusion Matrix on Validation Dataset for the Final Model, showing minimal error.}
    \label{fig:conf_matrix_test}
\end{figure}

\begin{figure}[t]
    \centering
    \includegraphics[width=14.0cm]{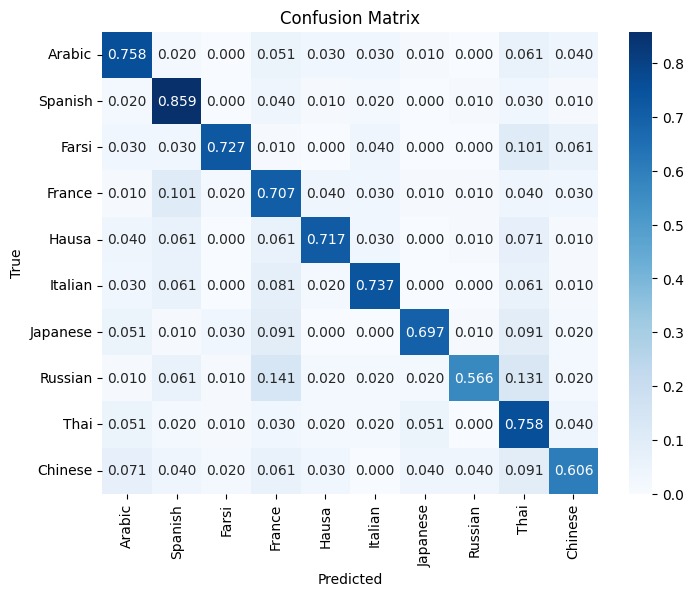}
    \caption{Confusion Matrix on Test Dataset for the Final Model, indicating potential overfitting compared to validation results.}
    \label{fig:conf_matrix_val}
\end{figure}

The results clearly demonstrate the efficacy of integrating 1x1 intermediate layers into the TDNN framework. This enhancement likely contributes positively, as detailed in the relevant section, by facilitating advanced data processing and generating more intricate feature patterns. Nonetheless, this increased model complexity raises concerns about potential overfitting, as the model might "memorize" data or inadvertently focus on identifying speakers.

A pivotal alteration was the transition to a funnel-shaped neuron configuration within the TDNN layers. Our hypothesis posits that the structural benefits of this configuration significantly boosted model performance. However, akin to the previous point, the augmented neuron count in this structure might also enable the model to more effectively memorize data or recognize speakers.

Another notable modification was the addition of 1x1 TDNN intermediate layers. The empirical data suggest substantial improvements from this integration. Future research should aim to identify the optimal number and arrangement of these intermediate layers.

In this study, the final results indicated that Spanish audio segments consistently outperformed Russian segments in both test and validation metrics. This disparity may be attributed to differences in segment lengths and the more extensive validation dataset for Spanish compared to Russian.

\section{Discussion}

Advancements in language recognition technologies have been noteworthy, yet they confront several challenges and limitations. This discussion highlights some of the prevalent issues:

\begin{enumerate}
    
    \item \textbf{Linguistic Ambiguity.} Distinguishing between languages with shared linguistic traits or diverse dialects remains challenging. For instance, accurately differentiating languages like Norwegian, Swedish, and Danish is complicated by their close linguistic similarities.

    \item \textbf{Code-Switching Phenomenon.} In multilingual settings~\cite{anidjar2023speech}, individuals often switch between languages in conversation or text~\cite{anidjar2021hybrid, anidjar2021thousand, anidjar2020thousand}, known as code-switching. This practice poses a challenge for language recognition systems, which may fail to correctly identify the languages involved, leading to misclassification or errors.

    \item \textbf{Scarcity of Diverse Training Data.} The efficacy of language recognition systems is contingent upon extensive and varied training datasets. The dearth of such data, especially for less prevalent or low-resource languages~\cite{anidjar2023crossing}, limits the system's performance, as comprehensive datasets are pivotal for accurate recognition.

    \item \textbf{Challenges with Uncommon or Emerging Languages.} While systems show proficiency with widely spoken languages, they often fall short in identifying less common or nascent languages due to insufficient data and resources.

    \item \textbf{Impact of Noise and Audio Quality.} Speech-based language recognition systems are susceptible to interference from background noise, poor audio quality, and diverse accents, impacting the accuracy of automatic speech recognition and subsequent language identification.

    \item \textbf{Constraints of Limited Context.} Language recognition systems relying on brief text or speech segments may lack adequate context~\cite{casakin2024use, casakindata} for accurate identification. This constraint is particularly problematic in differentiating between closely related languages, where context plays a crucial role.
    
\end{enumerate}

\section{Conclusions and Future Work}

In this study, we have enhanced a spoken language recognition model based on x-vectors, focusing on capturing long-term language features through a temporal pooling layer. Utilizing datasets from Common Voice, a premier repository for multilingual speech data, our work concentrated on ten languages from the Indo-European, Semitic, and East Asian linguistic families. We restructured the TDNN layers by incorporating 1X1 TDNN intermediate layers and transitioning into a funnel architecture. A comprehensive grid search was conducted to identify the optimal dilation and context size in the TDNN layers, enabling more effective detection of complex language patterns in audio segments. The model, after rigorous training and fine-tuning with augmented data, achieved an impressive accuracy of 97\%.

For future research, one promising avenue is the exploration of unverified data from Common Voice. While the current study utilized verified data, working with unverified or "dirty" data, characterized by inaccuracies and inconsistencies, can provide a more realistic representation of real-world scenarios. Additionally, cleaning 'dirty' audio files by removing non-essential parts could streamline the data, leaving only the most relevant segments for analysis. Another significant area for future work is the challenge of speaker recognition and separation in environments with overlapping speech. Developing methods to effectively distinguish individual speakers in such contexts would significantly enhance the model's applicability in real-life situations.

\bibliographystyle{unsrt}
\bibliography{main_new}
        
\end{document}